\documentclass[useAMS,usenatbib,usegraphicx]{mn2e}

\usepackage{times}

\newcommand{\gps}{\ensuremath{g_{\rm P1}}}
\newcommand{\rps}{\ensuremath{r_{\rm P1}}}
\newcommand{\ips}{\ensuremath{i_{\rm P1}}}
\newcommand{\zps}{\ensuremath{z_{\rm P1}}}
\newcommand{\yps}{\ensuremath{y_{\rm P1}}}
\newcommand{\grizy}{\gps\rps\ips\zps\yps}
\newcommand{\PS}{\protect \hbox {Pan-STARRS1}}

\title[A Thin Stellar Stream in Ophiuchus]{Serendipitous Discovery of a
         Thin Stellar Stream near the Galactic Bulge in the \PS\ 3$\pi$ Survey}
\author[E.~J.\ Bernard et al.]{%
Edouard J. Bernard,$^{1}$\thanks{E-mail: ejb@roe.ac.uk}
Annette M. N. Ferguson,$^{1}$
Edward F. Schlafly,$^{2}$
\newauthor
Mohamad Abbas,$^{3}$
Eric F. Bell,$^{4}$
Niall R. Deacon,$^{2}$
Nicolas F. Martin,$^{5,2}$
\newauthor
Hans-Walter Rix,$^{2}$
Branimir Sesar,$^{2}$
Colin T. Slater,$^{4}$
Jorge Pe\~narrubia,$^{1}$
\newauthor
Rosemary F. G. Wyse,$^{6}$
William S. Burgett,$^{7}$
Kenneth C. Chambers,$^{7}$
\newauthor
Peter W. Draper,$^{8}$
Klaus W. Hodapp,$^{7}$
Nicholas Kaiser,$^{7}$
Rolf-Peter Kudritzki,$^{7}$
\newauthor
Eugene A. Magnier,$^{7}$
Nigel Metcalfe,$^{8}$
Jeffrey S. Morgan,$^{7}$
Paul A. Price,$^{7}$
\newauthor
John L. Tonry,$^{7}$
Richard J. Wainscoat,$^{7}$
Christopher Waters$^{7}$\\
$^{1}$SUPA, Institute for Astronomy, University of Edinburgh, Royal
   Observatory, Blackford Hill, Edinburgh EH9 3HJ, UK \\
$^{2}$Max-Planck-Institut f\"ur Astronomie, K\"onigstuhl 17, D-69117
   Heidelberg, Germany \\
$^{3}$Astronomisches Rechen-Institut, Zentrum f\"ur Astronomie der
   Universit\"at Heidelberg, M\"onchhofstr. 12--14, D-69120 Heidelberg, Germany \\
$^{4}$Department of Astronomy, University of Michigan, 500 Church St.,
   Ann Arbor, MI 48109, USA \\
$^{5}$Observatoire Astronomique de Strasbourg, Universit\'e de Strasbourg,
   CNRS, UMR 7550, 11 rue de l'Universit\'e, F-67000 Strasbourg, France \\
$^{6}$Department of Physics and Astronomy, The Johns Hopkins University,
   3400 North Charles Street, Baltimore, MD 21218, USA \\
$^{7}$Institute for Astronomy, University of Hawaii, 2680 Woodlawn Drive,
   Honolulu HI 96822, USA \\
$^{8}$Department of Physics, Durham University, South Road, Durham DH1 3LE, UK
}

\begin{document}

\date{Accepted --. Received --; in original form --}

\pagerange{\pageref{firstpage}--\pageref{lastpage}} \pubyear{2014}

\maketitle

\label{firstpage}

\begin{abstract}

 We report the discovery of a thin stellar stream found in Pan-STARRS1
 photometry near the Galactic bulge in the constellation of Ophiuchus. It
 appears as a coherent structure in the colour-selected stellar density maps
 produced to search for tidal debris around nearby globular clusters. The
 stream is exceptionally short and narrow; it is about 2.5$\degr$ long and
 6$\arcmin$ wide in projection.
 The colour-magnitude diagram of this object, which harbours a blue
 horizontal-branch, is consistent with an old and relatively metal-poor
 population ([Fe/H]~$\sim-$1.3) located 9.5~$\pm$~0.9~kpc away at
 $(l,b) \sim (5\degr,+32\degr)$, and 5.0~$\pm$~1.0~kpc from the Galactic
 centre. These properties argue for a globular cluster as progenitor.
 The finding of such a prominent, nearby stream suggests that many streams
 could await discovery in the more densely populated regions of our Galaxy.

\end{abstract}


\section{Introduction}

 Ever since the discovery of massive tidal tails emerging from the globular
 cluster Palomar\,5 \citep{ode01}, the Milky Way field population has been
 the subject of intense scrutiny to find more of these cold stellar streams.
 The main incentives for finding streams are their potential use to constrain
 the shape and mass of the Milky Way dark matter halo
 \citep[e.g.][]{joh05,kop10,var11,pen12,ver13,lux13,deg14}, their ability to
 constrain the existence of mini-halos \citep{iba02,sie08,yoo11,car12}, as
 well as the fossil record they provide of the mass assembly history of the
 Milky Way.

\begin{figure*}
\includegraphics[width=13.5cm]{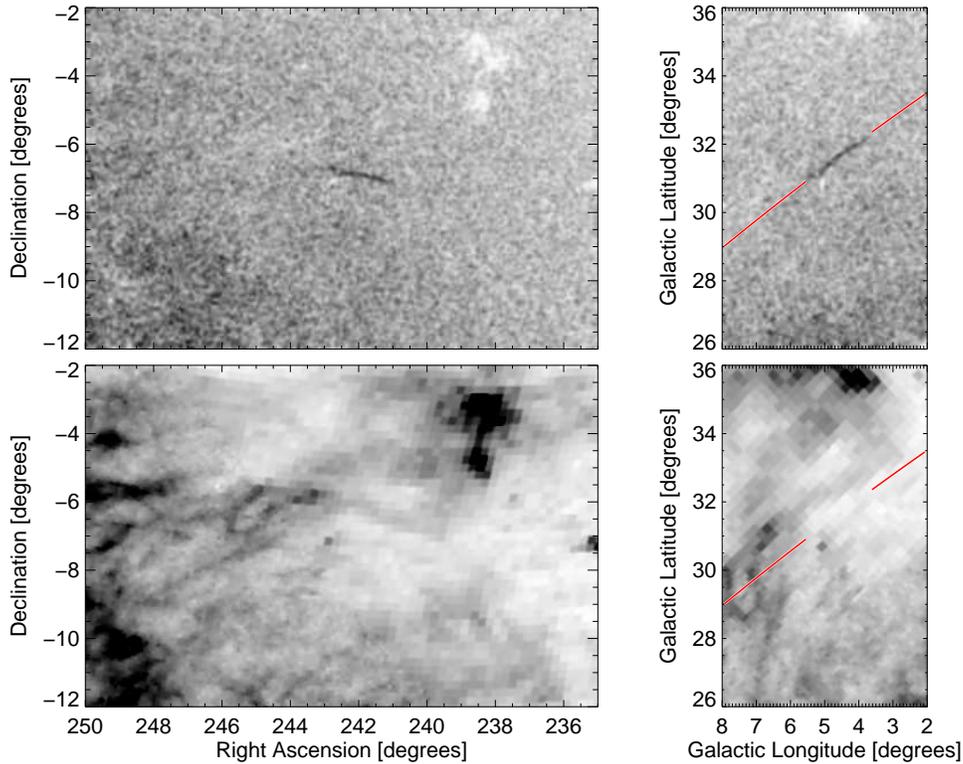}
\caption{{\it Top}: Density of stars with dereddened colours and magnitudes
 consistent with the main-sequence turn-off of an old and metal-poor
 population at heliocentric distances of 8--12~kpc (see selection box
 in Figure~\ref{fig:cmd}), shown in equatorial ({\it left}) and Galactic
 ({\it right}) coordinates. Darker areas indicate higher stellar density.
 {\it Bottom}: Reddening maps of the corresponding fields, derived from
 Pan-STARRS1 stellar photometry \citep[see][]{sch14}. The colour scale
 is logarithmic; white (black) corresponds to E(B$-$V)~=~0.17 (0.58).
 The stellar density and reddening maps have been smoothed
 with a gaussian kernel with full-width at half maximum of 12$\arcmin$ and
 6$\arcmin$, respectively. The stream is located close to the centre of each
 panel. The thick lines in the right panels trace the best-fitting great circle
 containing the stream.}
\label{fig:map}
\end{figure*}

 While a few streams have been revealed by kinematics alone
 \citep[e.g.][]{hel99,new09,wil11}, the vast majority were found by searching
 for coherent stellar over-densities in the homogeneous, wide-field photometric
 catalogue provided by the {\it Sloan Digital Sky Survey} \citep[SDSS;][]{yor00}
 \citep[e.g.][]{gri06a,gri06b,gri06c,gri06d,gri09,gri12,gri13,bel06,bel07,bon12},
 and more recently in VST ATLAS \citep{kop14}.
 As a result of SDSS's predominant high-latitude, Northern hemisphere coverage,
 the known streams are located far from the Galactic disc and bulge that were
 mostly avoided by these surveys.

 The \PS\ \citep[PS1;][]{kai10} 3$\pi$ Survey, observing the whole sky visible
 from Hawaii, has the advantage of providing some of the first deep imaging
 of the dense inner regions of our Galaxy. With a sky coverage spanning twice
 that of the SDSS footprint, it offers the possibility to survey these less
 well-studied areas to seek new tidal streams as well as further extensions
 of already known ones.

 In this paper, we report the discovery of a very thin stellar stream located
 in the inner Galaxy close to the Galactic bulge. It was found serendipitously
 when analysing PS1 data of wide areas around nearby globular clusters for
 the presence of tidal debris.
 We briefly describe the PS1 survey in Section~\ref{ps1}, and present the
 new stream and measurements of its properties in Section~\ref{phot}. A
 summary is given in Section~\ref{summ}.

\section{The \PS\ 3$\pi$ Survey}\label{ps1}

 The PS1 3$\pi$ Survey \citep[K.\,C.\ Chambers et al., in preparation]{kai10}
 is being carried out with the 1.8~m optical telescope installed on the peak of
 Haleakala in Hawaii. Thanks to the 1.4-Gigapixel imager \citep{ona08,ton09}
 covering a 7 square degree field-of-view ($\sim3.3\degr$ diameter), it is
 observing the whole sky north of $\delta>-30 \degr$ in five optical to
 near-infrared bands \citep[\grizy;][]{ton12b} up to four times per year.
 The exposure time ranges from 30 to 45 seconds, leading to median 5$\sigma$
 limiting AB magnitudes of 21.9, 21.8, 21.5, 20.7, and 19.7 for individual
 exposures in the \grizy bands, respectively \citep{mor12}. At the end of
 the survey, the 12 or so images per band will be stacked, increasing the
 depth of the final photometry by $\sim$1.2~mag \citep{met13}.

 The individual frames are automatically processed with the Image Processing
 Pipeline \citep{mag06} to produce a photometrically and astrometrically
 calibrated catalogue. A detailed description of the general PS1 data
 processing is given in \citet{ton12a}. The analysis presented in this paper
 is based on the photometric catalogue obtained by averaging the magnitudes
 of objects detected in individual exposures \citep{sch12}. At the end of the
 survey, the point source catalogue will be based on detections in the stacked
 images, leading to a significantly deeper photometry and better constraints
 on the tidal streams.

 The catalogue used here was first corrected for foreground reddening by
 interpolating the extinction at the position of each source using the
 \citet{sch14} dust maps with the extinction coefficients of \citet{sch11}.
 In this part of the sky, E(B$-$V) ranges from 0.17 to $\sim$1.2 (see
 Figure~\ref{fig:map}).
 We then cleaned the catalogue by rejecting non-stellar objects using the
 difference between PSF and aperture magnitudes, as well as poorly measured
 stars by keeping only objects with a signal-to-noise ratio of 10 or higher.

\begin{figure}
\includegraphics[width=8.5cm]{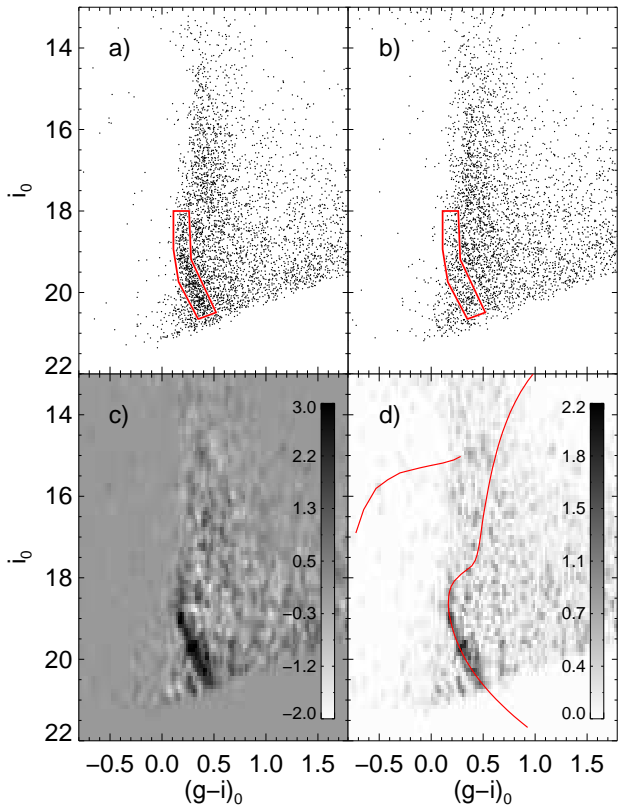}
\caption{Extinction corrected CMDs of the stream (a), a nearby field (b),
 the difference between the two (c), and the significance (d). The MSTO
 selection box used to produce Figure~\ref{fig:map} is shown by a red line in
 the top panels. The red lines in the last panel show the fiducial and HB of
 NGC\,5904 (M\,5) calculated from the PS1 photometry by \citet{ber14}. A few
 blue HB stars are visible in panel (a) at ($\gps-\ips$)$_0\sim-$0.5 and
 $i_{P1,0}\sim$~16.}
\label{fig:cmd}
\end{figure}

\section{A New Stream in Ophiuchus}\label{phot}

 The stream appears as a coherent structure in the maps showing the stellar
 density of objects with colours and magnitudes corresponding to the old,
 metal-poor main-sequence turn-off (MSTO) of nearby globular clusters. The
 colour and magnitude cuts were then refined based on the distribution of
 stars in the colour-magnitude diagram (CMD) of the stream region (see below).
 The resulting map is shown
 in Figure~\ref{fig:map} in both equatorial and Galactic coordinates, along
 with the corresponding reddening maps from \citet{sch14}. While there is a
 little residual substructure due to the strong differential reddening in
 this part of the sky, the stream is obvious at $\alpha \sim 242\degr$ and
 $\delta \sim -7\degr$ ($l\sim 5\degr$, $b\sim +32\degr$). There are no
 features in the dust map that could produce such an artifact.
 In fact, we checked that we recover the stream whether or not we apply the
 reddening correction.

 Figure~\ref{fig:cmd} shows the CMD of a 9$\arcmin$ wide box centred on the
 stream and extending between 241\fdg5~$<\alpha<$~243\fdg5, as well as a nearby
 comparison region of the same area. The bottom panels show the residuals and
 the significance of the residuals in Poissonian sigmas. The MSTO is visible
 as an over-density at ($\gps-\ips$)$_0\sim$~0.2 and $i_{P1,0}\sim$~19, and is
 highlighted in the top panels by a red box. A few blue
 horizontal-branch (HB) stars are also detected at ($\gps-\ips$)$_0\sim-$0.5 and
 $i_{P1,0}\sim$~16. As expected from such a sparse HB, no RR~Lyrae star could be
 found at the location and distance of the stream in the catalogue of
 \citet{dra13}. The lack of obvious sub-giant and red giant branches does not
 allow tight constraints on the constituent stellar populations, although the
 presence of a blue HB suggests an old and metal-poor population ($\ga$~10~Gyr
 old, [Fe/H]~$\la-$~1.0). The features of the stream CMD are well fitted
 by the fiducial of the old globular cluster NGC\,5904 (M\,5, [Fe/H]~$\sim-$1.3)
 shown as red lines \citep[from][]{ber14}. The fiducial was first corrected for
 the reddening along the line of sight to NGC\,5904
 \citep[E(B$-$V)~$\sim$~0.09;][]{sch14}, then shifted by $+$0.5~mag to match
 the luminosity of the stream MSTO. A small colour shift (0.07~mag to the blue)
 improves the fit to the MSTO, possibly due to the high and varying reddening
 in this region. The HB of the fiducial also provides a good fit to the
 observed blue HB stars, giving further support that the estimated distance is
 reliable. Assuming (m$-$M)$_0$~=~14.44$\pm$0.02 for NGC\,5904 \citep{cop11}
 yields a true distance modulus of (m$-$M)$_0$~=~14.9$\pm$0.2
 (i.e.\ 9.5$\pm$0.9~kpc) for the Ophiuchus stream. Finally, if we assume
 that the Sun is located 8~kpc from the Galactic centre (GC), we find a
 Sun--GC--stream angle of $\sim$89 degrees, placing the stream almost directly
 above the Galactic bulge at a Galactocentric distance of 5.0~$\pm$~1.0~kpc.

\begin{table}
 \begin{minipage}{80mm}
\centering
\caption{Summary of the Stream Properties.}
\label{tab:prop}
\begin{tabular}{lc}
\hline
 Parameter               & Value                                          \\
\hline
 R.A.\ (J2000.0)         & \ \ 16:07:12                                   \\
 Dec.\ (J2000.0)         &  $-$06:55:30                                   \\
 $l$                     & \ 4\fdg53                                      \\
 $b$                     & +31\fdg69                                      \\
 (m$-$M)$_0$             & 14.9~$\pm$~0.2                                 \\
 Median E(B$-$V)         & 0.23                                           \\
 Heliocentric distance   & 9.5~$\pm$~0.9~kpc                              \\
 Galactocentric distance & 5.0~$\pm$~1.0~kpc                              \\
 Width (FWHM)            & 7.0$\arcmin \ \pm$~0.8$\arcmin$ (19~$\pm$~2~pc)\\
 Length                  & $\sim 2.5\degr$ ($\sim 400$~pc)                \\
 $M_V$                   & $-3.0\pm0.5$                                   \\
 $L_V$                   & $1.4 \pm 0.6 \times 10^3 L_{\sun}$             \\
\hline
\end{tabular}
\end{minipage}
\end{table}

 To estimate the total luminosity of the stream, we used {\sc IAC-Star}
 \citep{apa04} to generate the CMD of an old population (11.5--12.5~Gyr)
 containing 10$^5$ stars with a metallicity comparable to that of NGC\,5904
 and with the same depth as our observations -- stars fainter than our
 detection limit have a negligible contribution to the total magnitude.
 In Figure~\ref{fig:cmd}, we find that there are about 500 more stars in
 panel (a) containing the stream than in the comparison field (panel (b)).
 Summing the luminosity of a sample of 500 stars extracted randomly from the
 synthetic CMD gives the total flux. We repeated this step 10$^4$ times,
 extracting between 300 and 700 stars each time, and found a total
 magnitude and luminosity of $M_V=-3.0\pm0.5$ and
 $L_V=1.4 \pm 0.6 \times 10^3 L_{\sun}$, respectively.
 These are comparable to some of the fainter halo clusters (e.g.\ Whiting\,1,
 Terzan\,9, Palomar\,1, Palomar\,13: \citealt[2010 Edition]{har96}),
 although a significant fraction of the stars may have already been stripped
 off and lie further out along the stream orbit.

\begin{figure*}
\includegraphics[width=13.5cm]{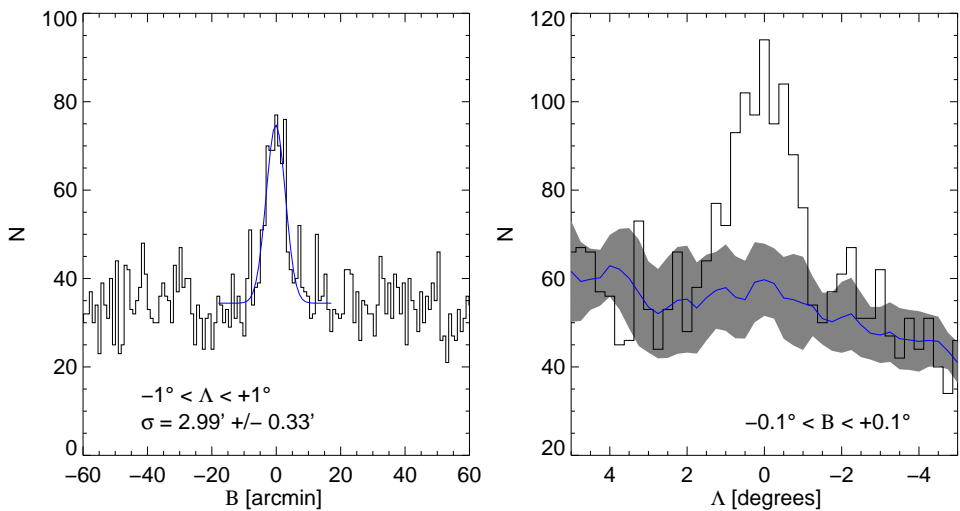}
\caption{{\it Left}: MSTO star density as a function of stream latitude.
 The stream is detected as a strong over-density that is well fitted
 by a gaussian with $\sigma$~=~2.99$\arcmin\pm$0.33$\arcmin$.
 {\it Right}: Stellar density as a function of stream longitude. The blue
 line and shaded area represent the mean background contamination and its
 dispersion (see text).}
\label{fig:lenwid}
\end{figure*}

 The extent of the stream on the sky was calculated by reprojecting the
 stellar maps to a new spherical coordinate system ($\Lambda$,$B$) with
 the pole located at ($\alpha$,$\delta$) = (184\fdg32, +77\fdg25)
 [($l$,$b$) = (125\fdg37, +39\fdg72)]. In this system the stream approximately
 lies along the equator -- shown by the red lines in Figure~\ref{fig:map} --
 making it easier to measure its width and length. In Figure~\ref{fig:lenwid},
 we show the distribution of MSTO stars across the $\Lambda$ and $B$
 dimensions. The left panel shows the stellar density across the stream,
 where all the MSTO stars in the range $-1\degr<\Lambda<1\degr$ have been
 used. The stream is detected as a significant over-density at $B\sim0\degr$.
 The profile is best fit by a gaussian with $\sigma$~=
 2.99$\arcmin\pm$0.33$\arcmin$, corresponding to a full-width at half maximum
 (FWHM) of 7.0$\arcmin\pm$0.8$\arcmin$, i.e.\ 19~$\pm$~2~pc at the distance of
 the stream. However, we note that in Figure~\ref{fig:map} the stream appears
 slightly more curved than the great circle shown, implying that the
 intrinsic width may be even smaller. The right panel shows the profile along
 the length of the stream, for which we used the
 MSTO stars in a narrow strip (12$\arcmin$) centred on the stream. To estimate
 the background, we selected MSTO stars in 6 identical strips at higher and
 lower $B$; their average and dispersion are shown as the blue line and
 the shaded area. The on-stream histogram shows a significant over-density
 between $-1\fdg2\la\Lambda\la1\fdg3$, and is therefore about 2\fdg5
 long (i.e.\ $\sim$400~pc in projection).

 Table~\ref{tab:prop} summarizes the estimated properties of the stream:
 we list the approximate coordinates of its centre, the heliocentric and
 Galactocentric distances, the extent on the sky, and the estimated luminosity.

\section{Discussion and Conclusions}\label{summ}

 We have identified a new stellar stream in the constellation of Ophiuchus,
 a part of the sky that has rarely been searched for streams because of the
 high stellar density and significant differential and foreground reddening.
 Both the morphology of the MSTO and the presence of a blue HB are typical
 of an old and metal-poor population ($\ga$~10~Gyr old, [Fe/H]~$\la-$~1.0).
 These properties, along with the small width of the stream and absolute
 magnitude suggest a globular cluster as progenitor.

 We find that the stream is exceptionally short ($\sim$2\fdg5,
 i.e.\ $\sim$400~pc, in projection) compared
 to all the other streams found so far, that are usually several tens of
 degrees long. For comparison, the shortest stream known to date is the
 Pisces--Triangulum stream, which has been traced over $\sim$15\degr
 (i.e.\ $\sim$7~kpc) on the sky \citep[e.g.][]{bon12,mar14}.
 We have explored shifting the MSTO selection box in magnitude to account
 for possible effects of differential reddening residuals and extension
 of the stream along the line of sight, but failed to detect any over-density
 beyond the current extent. This experiment did reveal a possible distance
 gradient along the stream, with the eastern tail being closer to the Sun,
 although the apparent change in MSTO magnitude could simply be a consequence
 of the differential reddening.

 Surprisingly, neither the stellar density along the stream nor a careful
 visual inspection of the images reveals a potential remnant of the
 progenitor. This suggests that it has already been completely disrupted.
 Given that the length of a tidal stream is a function of the time since the
 stars became unbound, a short stream may indicate that the progenitor has
 been disrupted only recently. However, this scenario is hard to reconcile
 with the lack of an obvious progenitor in the vicinity of the stream.
 Another possibility is that we are observing the stars of a fully disrupted
 cluster at apogalacticon on a highly elliptical orbit: at this point of the
 orbit, unbound stars tend to clump together because of the slower orbital
 velocity \citep[e.g.][]{deh04,kup12}. The data currently available are not
 sufficient to reliably trace the orbit of the progenitor, which would help us
 understand its past evolution and likely fate; radial velocities and
 proper motions of a sample of stream members will be crucial for this purpose.

 Spectroscopic metallicities of stream stars will also shed light on the
 nature of the progenitor, and help understand how such events may contribute
 to the stellar populations in the central regions of the Galaxy.
 For example, recent spectroscopic surveys of the outer Galactic bulge have
 revealed the presence of a significant number of metal-poor stars
 ([Fe/H]~$<-$1; e.g.\ \citealt{gon11,gar13}). Their origin may be linked to
 tidal stripping events such as the one we are witnessing with this stream.

\section*{Acknowledgments}

E.J.B.\ and A.M.N.F.\ are grateful to Douglas C.\ Heggie and Anna~Lisa Varri
for enlightening discussions. The authors would like to thank the anonymous
referee for a prompt report and useful comments. This research was supported
by a consolidated grant from the Science Technology and Facilities Council.
E.F.S. and N.F.M. acknowledge support from the DFG's grant SFB881 (A3) ``The
Milky Way System". N.F.M. gratefully acknowledges the CNRS for support through
PICS project PICS06183. The research leading to these results has received
funding from the European Research Council under the European Union's Seventh
Framework Programme (FP 7) ERC Grant Agreement n.\ [321035].

The PS1 Surveys have been made possible through contributions of
the Institute for Astronomy, the University of Hawaii, the Pan-STARRS Project
Office, the Max-Planck Society and its participating institutes, the Max Planck
Institute for Astronomy, Heidelberg and the Max Planck Institute for
Extraterrestrial Physics, Garching, The Johns Hopkins University, Durham
University, the University of Edinburgh, Queen's University Belfast, the
Harvard-Smithsonian Center for Astrophysics, the Las Cumbres Observatory Global
Telescope Network Incorporated, the National Central University of Taiwan, the
Space Telescope Science Institute, the National Aeronautics and Space
Administration under Grant No.\ NNX08AR22G issued through the Planetary Science
Division of the NASA Science Mission Directorate,  the National Science
Foundation under Grant No.\ AST-1238877, and the University of Maryland.

This work has made use of the IAC-STAR Synthetic CMD computation code.
IAC-STAR is supported and maintained by the computer division of the Instituto
de Astrof\'isica de Canarias.

\label{lastpage}

\bsp

\end{document}